\documentclass[aps,pre,reprint,superscriptaddress,amsmath,showpacs]{revtex4-1}
\usepackage[breaklinks,colorlinks=true]{hyperref}
\usepackage{graphicx}
\usepackage{amssymb}

\begin{document}

\title{Effects of junctional correlations in the totally asymmetric simple exclusion process on random regular networks}

\author{Yongjoo Baek}
\email[Present address: Department of Physics, Technion--Israel Institute of Technology, Haifa 32000, Israel] {}
\affiliation{Natural Science Research Institute,
Korea Advanced Institute of Science and Technology, Daejeon
305-701, Korea}

\author{Meesoon Ha}
\email[Corresponding author: ]{msha@chosun.ac.kr}
\affiliation{Department of Physics Education, Chosun University,
Gwangju 501-759, Korea}

\author{Hawoong Jeong}
\email[ ]{hjeong@kaist.edu}
\affiliation{Department of Physics and Institute for the
BioCentury, Korea Advanced Institute of Science and Technology,
Daejeon 305-701, Korea}
\affiliation{APCTP, Pohang, Gyeongbuk
790-784, Korea}

\date{\today}

\begin{abstract}
We investigate the totally asymmetric simple exclusion process on
closed and directed random regular networks, which is a simple
model of active transport in the one-dimensional segments coupled
by junctions. By a pair mean-field theory and detailed numerical
analyses, it is found that the correlations at junctions induce
two notable deviations from the simple mean-field theory which
neglects these correlations: (1) the narrower range of particle
density for phase coexistence and (2) the algebraic decay of
density profile with exponent $1/2$ even outside the
maximal-current phase. We show that these anomalies are
attributable to the effective slow bonds formed by the network
junctions.
\end{abstract}



\maketitle

\section{Introduction} \label{sec:intro}

Various transport phenomena involve self-driven, hard-core particles moving unidirectionally along one-dimensional (1D) segments. The {\em totally asymmetric simple exclusion process} (TASEP), in which 1D lattice gas particles randomly hop one step forward if and only if the next site is vacant, is one of the simplest models of such phenomena. The model has an important advantage of being exactly solvable in homogeneous 1D systems~\cite{Derrida1993,*Schutz1993}, with very well-understood dynamical phases~\cite{Blythe2007,*Derrida2007}. Besides its original purpose as the model of mRNA translation by ribosomes~\cite{MacDonald1968}, the TASEP has been applied in modified forms to numerous examples of vehicular, pedestrian, and biological transport~\cite{Helbing2001,*Chowdhury2000,*Chowdhury2005,*Bressloff2013}.

In many of these examples, the 1D segments are not isolated from each other but coupled by junctions, where the particles can randomly switch from one segment to another. The motor proteins on cytoskeletal networks~\cite{Alberts2007,Ali2007} provide one interesting example of such behavior. In order to clarify the effects of junctions, the TASEP has been studied in various systems consisting of coupled 1D segments, including {\em open} (i.e. connected to particle reservoirs) segments with a single junction~\cite{Brankov2004,*Pronina2005,*Pesheva2013}, {\em closed} (i.e. conserving the particles) 1D loops with a single junction~\cite{Embley2009,Raguin2013}, closed 1D loops with a shortcut~\cite{Yuan2007,*Bunzarova2014}, periodic hexagonal lattices~\cite{Embley2008}, and closed and directed random regular (CDRR) networks~\cite{Neri2011}. For lack of exact solutions, all these studies rely on the simplifying assumption that the correlations between each junction and its neighboring sites are negligible. This approach is also called the defect mean-field (DMF) theory~\cite{Foulaadvand2008}, because it was originally proposed to address the effects of a single local defect in the 1D TASEP~\cite{Janowsky1992,*Janowsky1994,Kolomeisky1998a}. The theory allows one to approximate every segment as an open 1D system whose boundary conditions are given by occupancies of the junctions at both ends. Thus, in the limit of infinitely long segments, the dynamical phases of different parts of the system can be analytically predicted from the well-established knowledge of the 1D systems.

While the predictions of the DMF theory agree qualitatively well
with the numerical results, they are also known to be
quantitatively inexact for the single-junction
cases~\cite{Kolomeisky1998a,Embley2009} due to the neglected
correlations at junctions. When the system has a large number of
junctions, to our knowledge there has been no systematic test of
the quantitative agreement in the proper asymptotic limit. Using
the CDRR networks studied by the authors of Ref.~\cite{Neri2011},
in this study we analytically and numerically show that the
neglected junctional correlations induce nonzero corrections to
the DMF predictions. On the analytical side, we develop a defect
pair mean-field (DPMF) theory which takes into account the pair
correlations between junctions and their neighboring sites. On the
numerical side, we detect non-DMF behaviors in the steady-state
properties of the current and the density profile of particles
through extensive Monte Carlo simulations at different segment
lengths. The observed non-DMF behaviors reveal interesting
connections to the unresolved issues of how a {\em single slow
bond} affects the 1D
transport~\cite{Janowsky1992,*Janowsky1994,Tang1993,*Balents1994,*Krug1994,*Kinzelbach1995,*Hwa1995,*Lassig1998,Kolomeisky1998a,MHa2003,JHLee2009}.

The rest of the paper is organized as follows. We first introduce the model in Sec.~\ref{sec:model} and explain its DMF description in Sec.~\ref{sec:DMF}. Then we show by a DPMF argument that there are nonzero corrections to this DMF description in Sec.~\ref{sec:DPMF}. Numerical evidence for these corrections is presented in Sec.~\ref{sec:results}, whose similarities with the 1D TASEP with a single slow bond are discussed in Sec.~\ref{sec:SB}. Finally, we summarize our results and conclude in Sec.~\ref{sec:conclusions}.

\begin{figure*}
    \centering
    \includegraphics[width=0.8\textwidth]{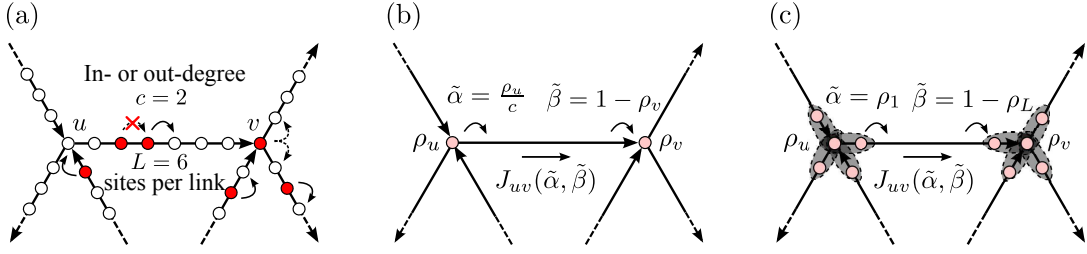}
    \caption{\label{fig:model} (Color online) TASEP on the CDRR networks and its approximate descriptions. (a) Each junction (node) has $c$ outgoing and $c$ incoming links, where each link is by itself a chain of $L$ inner sites. The dynamics is identical to the 1D TASEP except that a particle at a junction randomly chooses one of the $c$ adjacent sites in the outgoing links as its next destination. (b) The DMF theory assumes that each link is an open 1D system whose entry rate $\tilde{\alpha}$ and exit rate $\tilde{\beta}$ are given by the occupancy $\rho_u$ of each junction. (c) The DPMF theory assumes that only the pairs consisting of a junction and its adjacent site are correlated. The entry (exit) rate of a link is given by the occupancy $\rho_1$ ($\rho_L$) of the first (last) inner site of the link.}
\end{figure*}

\section{Model} \label{sec:model}

We consider a closed and directed random regular (CDRR) network of $N$ junctions (nodes) connected to each other by directed links. Each junction has $c$ outgoing and $c$ incoming links, whose arrangements are completely random as long as every pair of junctions can reach one another by some path preserving the link directions. These conditions ensure that the system is closed and irreducible to disconnected subcomponents.

Now we replace each directed link between two junctions with a 1D chain of $L$ {\em inner sites} [see Fig.~\ref{fig:model}(a)], so that the system has $N_\mathrm{tot} \equiv cN(L+1)$ {\em sites} (junctions and inner sites) in total. Each site can hold at most a single particle, and the positions of particles are updated as follows:
\begin{enumerate}
    \item Randomly choose a site with equal probability.
    \item If the chosen site is an occupied inner site with a vacant neighbor (an inner site or a junction) in the link direction, then move the particle in the former site to the latter.
    \item If the chosen site is an occupied junction, randomly choose one of the $c$ outgoing links with equal probability. If the first inner site of the chosen outgoing link is vacant, then move the particle in the junction to the inner site.
\end{enumerate}
In other words, (1) the transport along each link is equivalent to the ordinary 1D TASEP, (2) the exclusion principle also applies to all junctions, and (3) each junction distributes the current to its outgoing links in an unbiased manner. Since the number of particles $M$ remains constant throughout the dynamics, the global particle density $\rho \equiv M/N_\mathrm{tot}$ becomes a control parameter. Each update of a site increases the time by $\Delta t = N_\mathrm{tot}^{-1}$ so that every site is updated once per unit time on average. The global current $J$ is defined as the average number of hops per unit time divided by the total number of sites.

\section{Defect mean-field theory} \label{sec:DMF}

The model described in the previous section can be approximately described by the defect mean-field (DMF) theory, which neglects the correlations between each junction and its adjacent sites. Let us consider a directed link from junction $u$ to junction $v$. The state of a junction (say $u$) is denoted by $\tau_u$, which is either $1$ (occupied) or $0$ (vacant). If $\langle X \rangle$ represents the ensemble average of an observable $X$, then the mean occupancy of junction $u$ can be written as $\rho_u \equiv \langle \tau_u \rangle$. According to the DMF theory, each link can be regarded as an open 1D system whose entry and exit rates are given by
\begin{align} \label{eq:DMF_eff_rates}
    \tilde{\alpha} = \rho_u/c, \quad
    \tilde{\beta} = 1 - \rho_v,
\end{align}
respectively [see Fig.~\ref{fig:model}(b)]. Then the steady-state current $J_{uv}(\tilde{\alpha},\tilde{\beta})$ and the bulk density $\rho_{uv}(\tilde{\alpha},\tilde{\beta})$ of link $uv$ can be obtained from the exact solutions of 1D TASEP with open boundaries~\cite{Derrida1993,*Schutz1993}.

\begin{figure}
    \centering
    \includegraphics[width=\columnwidth]{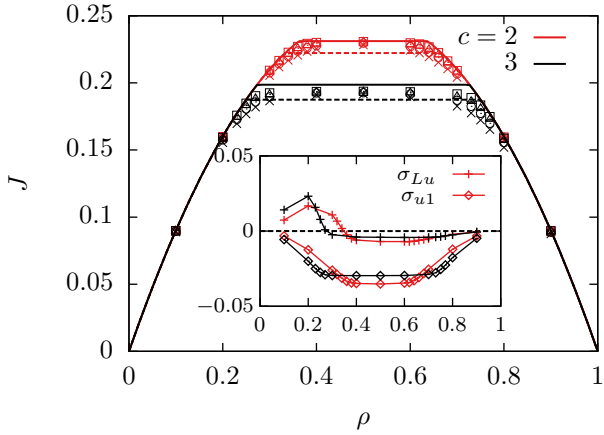}
    \caption{\label{fig:den_cur} (Color online) Fundamental diagram of the density--current relation for the CDRR networks with $c = 2$ (red/gray) and $c = 3$ (black). The DMF predictions (dashed lines), the DPMF predictions (solid lines), and the simulation results obtained at $L = 100$ ($\times$), $200$ ($\bigcirc$), $400$ ($\bigtriangleup$), $800$ ($\Box$) are shown. (Inset) Covariance ($\sigma_{ab} \equiv \langle \tau_a\tau_b \rangle - \langle \tau_a \rangle \langle \tau_b \rangle$) between a junction and its adjacent site in an incoming link (upper two curves) and in an outgoing link (lower two curves) at $L = 800$, where the lines are guide to the eye. The CDRR networks with $N = 10^3$ are used.}
\end{figure}

In the steady state, each junction satisfies the continuity equation
\begin{equation} \label{eq:DMF_continuity}
    \sum_v J_{uv} = \sum_{v'} J_{v'u},
\end{equation}
which is also called {\em Kirchhoff's current law} in the context of electrical circuits. This equation is automatically satisfied if all junctions have the same mean occupancy $\rho_u$, so that the link indices (e.g. $uv$) can be dropped from all quantities mentioned so far. Then the global current $J$ is equal to the current through each link, which gives
\begin{align}
J &= \min \left[\tilde{\alpha}(1-\tilde{\alpha}),\tilde{\beta}(1-\tilde{\beta})\right] \nonumber \\
&= \begin{cases}
(\rho_u/c) (1-\rho_u/c) &\text{ if $\rho_u < c/(c+1)$},\\
c/(c+1)^2 &\text{ if $\rho_u = c/(c+1)$},\\
\rho_u(1-\rho_u) &\text{ otherwise.}
\end{cases}
\end{align}
Similarly, the global particle density $\rho$ is equal to the bulk density of particles in each link, which satisfies
\begin{align}
\rho = \begin{cases}
\rho_u/c &\text{ if $\rho_u < c/(c+1)$},\\
\rho_u  &\text{ if $\rho_u > c/(c+1)$}.
\end{cases}
\end{align}
When $\rho_u = c/(c+1)$, $\rho$ can assume any value between $\rho_u/c$ and $\rho_u$, since each link has two coexistent blocks with different bulk densities whose interface can fluctuate back and forth~\cite{Kolomeisky1998b}. Hence the density--current relation is obtained as
\begin{equation} \label{eq:MF_cur}
J = \begin{cases}
c/(c+1)^2 &\text{ if $\rho_\mathrm{DMF}^* < \rho < 1-\rho_\mathrm{DMF}^*$}\\
\rho (1-\rho) &\text{ otherwise},
\end{cases}
\end{equation}
where $\rho_\mathrm{DMF}^* \equiv 1/(c+1)$ and $1 - \rho_\mathrm{DMF}^*$ become the phase boundaries. Here $\rho < \rho_\mathrm{DMF}^*$ corresponds to the low-density phase, $\rho > 1 - \rho_\mathrm{DMF}^*$ to the high-density phase, and the current plateau in the middle to the phase coexistence regime. The predictions of Eq.~(\ref{eq:MF_cur}) are shown by the dashed lines in the main plot of Fig.~\ref{fig:den_cur}.

\section{Defect pair mean-field theory} \label{sec:DPMF}

In this section we propose the defect pair mean-field (DPMF) theory, which takes into account the pair correlations between junctions and their adjacent sites. As in the DMF case, we expect all links to have the same statistics, which implies that the link indices can be left out. Thus we let $\tau_i$ denote the state of the $i$-th inner site from the entrance junction, whose mean occupancy is denoted by $\rho_i \equiv \langle \tau_i \rangle$.

The DPMF theory assumes that the only nonzero correlations among $\tau_L$, $\tau_u$ and $\tau_1$ are the pair correlations between $\tau_L$ and $\tau_u$ and those between $\tau_u$ and $\tau_1$. These correlations are related to the conditional probabilities
\begin{align} \label{eq:mu_def}
\mu^{\tau_L \tau_u}_u &\equiv \mathrm{P}(\tau_u|\tau_L), &\; \mu^{\tau_u \tau_1}_1 &\equiv \mathrm{P}(\tau_1|\tau_u), \nonumber \\
\nu^{\tau_u \tau_L}_L &\equiv \mathrm{P}(\tau_L|\tau_u), &\; \nu^{\tau_1 \tau_u}_u &\equiv \mathrm{P}(\tau_u|\tau_1).
\end{align}
For example, the three-point correlation $\langle \tau_L (1-\tau_u) \tau_1 \rangle$ can be written as
\begin{align} \label{eq:corr_ex}
\langle \tau_L (1-\tau_u) \tau_1 \rangle &= \mathrm{P}(\tau_L = 1, \tau_u = 0, \tau_1 = 1) \nonumber \\
&= \rho_L \mu_u^{10} \mathrm{P} (\tau_1 = 1 | \tau_u = 0, \tau_L = 1) \nonumber \\
&= \rho_L \mu_u^{10} \mu_1^{01},
\end{align}
where the third equality follows from the assumption that there is no correlation between $\tau_L$ and $\tau_1$. The theory also assumes that these $\tau_L$, $\tau_u$ and $\tau_1$ are uncorrelated with the rest of the system, so that the other inner sites of each link can be regarded as forming an open 1D system whose entry and exit rates are given by
\begin{align} \label{eq:DPMF_eff_rates}
    \tilde{\alpha} = \rho_1/c, \quad \tilde{\beta} = 1 - \rho_L,
\end{align}
respectively [see Fig.~\ref{fig:model}(c)].

The correlations between a junction and its adjacent sites evolve according to
\begin{align}
\frac{d}{dt}\langle \tau_u \tau_1 \rangle &= c \langle \tau_L (1-\tau_u) \tau_1 \rangle - \langle \tau_u \tau_1 (1-\tau_2)\rangle \nonumber \\
&\quad - (c-1) \langle \tau_u (1-\tau_1) \tau_1' \rangle, \nonumber \\
\frac{d}{dt}\langle \tau_L \tau_u \rangle &= c \langle \tau_{L-1} (1-\tau_L) \tau_u \rangle - \langle \tau_L \tau_u (1-\tau_1)\rangle \nonumber \\
&\quad + (c-1) \langle \tau_L \tau_L' (1-\tau_u) \rangle, \label{eq:DPMF}
\end{align}
where $\tau_1$ and $\tau_1'$ ($\tau_L$ and $\tau_L'$) denote the states of the first (last) inner sites belonging to different outgoing (incoming) links. In the steady state, through manipulations similar to Eq.~(\ref{eq:corr_ex}), these equations can be rewritten as
\begin{align}
c \rho_L \mu_u^{10} \mu_1^{01} &= \rho_u \mu_1^{11}(1-\rho_2) + (c-1) \rho_u \left(\mu_1^{10}\right)^2, \nonumber\\
c \rho_{L-1} (1-\rho_L) \mu_u^{01} &= \rho_L \mu_u^{11}\mu_1^{10} \nonumber \\
&\quad + (c-1) (1-\rho_u) \left(\nu_L^{01}\right)^2. \label{eq:DPMF_steady}
\end{align}
From the steady-state conditions of the single-site occupancies and the definitions of conditional probabilities given by Eq.~(\ref{eq:mu_def}), we obtain the following useful identities:
\begin{align}
J &= (1-\rho_u)\nu_{L}^{01} = \rho_{L}\mu_{u}^{10} = \rho_u\mu_1^{10}/c \nonumber \\
&= \rho_{L-1} (1-\rho_L) = \rho_1 (1-\rho_2), \nonumber \\
\mu_u^{11} &= 1-\mu_u^{10}, \quad \mu_1^{11} = 1-\mu_1^{10}, \nonumber \\
\mu_1^{01} &= \frac{\rho_1 - \rho_u + cJ}{1-\rho_u}, \quad \mu_u^{01} = \frac{\rho_u - \rho_L + J}{1-\rho_L}.
\end{align}
Using these identities, Eq.~(\ref{eq:DPMF_steady}) can be rewritten as
\begin{align}
\frac{c(\rho_1-\rho_u+cJ)}{1-\rho_u} &= \frac{\rho_u-cJ}{\rho_1} + \frac{c(c-1)(\rho_u-cJ)}{\rho_u}, \nonumber \\
\frac{\rho_u - \rho_L + J}{1-\rho_L} &= \frac{c(\rho_L - J)}{\rho_u} - \frac{(c-1)J}{1-\rho_u} \label{eq:DPMF_ss}.
\end{align}

At the phase boundary $\rho_\mathrm{DPMF}^*$ between the low-density phase and the coexistence regime, we expect that the sites adjacent to the entrance (exit) junction have the occupancy equal to the bulk density of the low-density (high-density) phase. In order to satisfy the steady-state condition, both bulk densities must correspond to the same value of current, which implies
\begin{equation} \label{eq:PA_phase_boundary}
\rho_1 = 1 - \rho_L = \rho_\mathrm{DPMF}^*, \quad J = \rho_\mathrm{DPMF}^*(1-\rho_\mathrm{DPMF}^*).
\end{equation}
Plugging this condition into Eq.~(\ref{eq:DPMF_ss}), we obtain a system of equations for two unknown variables $\rho_\mathrm{DPMF}^*$ and $\rho_u$. They can be solved for $\rho_\mathrm{DPMF}^*$ as
\begin{align} \label{eq:PA_correction}
\rho_\mathrm{DPMF}^* &= \frac{1+4c-\sqrt{1+4c+4c^2-4c^3+4c^4}}{2c(1+2c-c^2)} \nonumber \\
&\approx \rho_\mathrm{DMF}^* \left[1 + \frac{1}{2}\left(\frac{1}{c+1}\right) - \frac{1}{8}\left(\frac{1}{c+1}\right)^2 \right].
\end{align}
Thus, the DPMF theory suggests that junctional correlations induce nonzero corrections to the DMF predictions in the asymptotic limit: the phase coexistence regime spans a narrower range of $\rho$ and has a larger value of $J$ (see the solid lines in the main plot of Fig.~\ref{fig:den_cur}). This is reminiscent of the observation that the current plateau of the TASEP on closed loops with a single junction seems to be underestimated by the DMF prediction~\cite{Embley2009}. We numerically verify those corrections for the CDRR networks in the next section.

\section{Numerical results} \label{sec:results}

\begin{figure}
    \centering
    \includegraphics[width=\columnwidth]{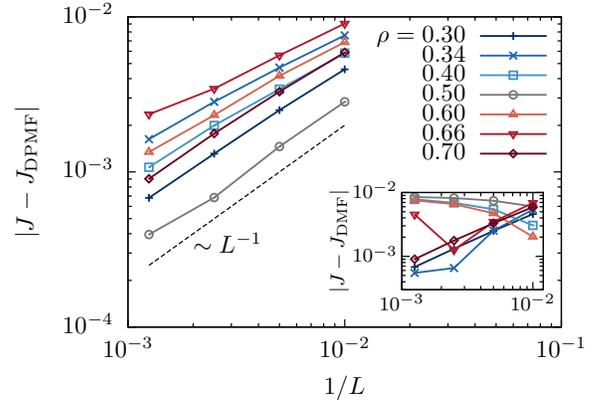}
    \caption{\label{fig:scaling} (Color online) Deviations of the simulation results from the DPMF (main plot) and the DMF (inset) predictions. The CDRR networks with $N = 10^3$ and $c = 2$ are used. The lines are guides to the eye.}
\end{figure}

\subsection{Density--current relation}

\begin{figure}
    \centering
    \includegraphics[width=0.965\columnwidth]{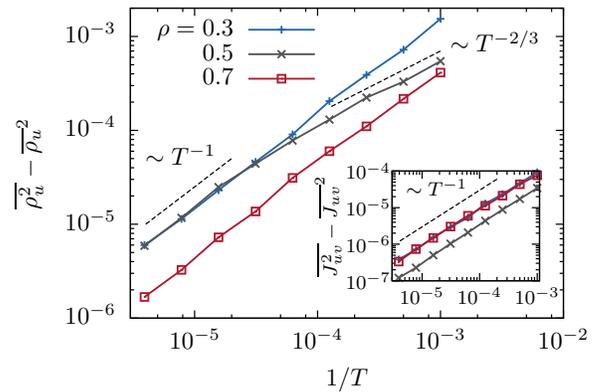}
    \caption{\label{fig:time_variance} (Color online) Variances of the mean occupancies $\rho_u$ among difference junctions (main plot) and the mean currents $J_{uv}$ among different links (inset) as functions of the observation time $T$. A CDRR network with $N = 10^3$, $c = 2$, and $L = 10^2$ is used. The lines are guide to the eye.}
\end{figure}

\begin{figure*}
    \centering
    \includegraphics[width=0.88\textwidth]{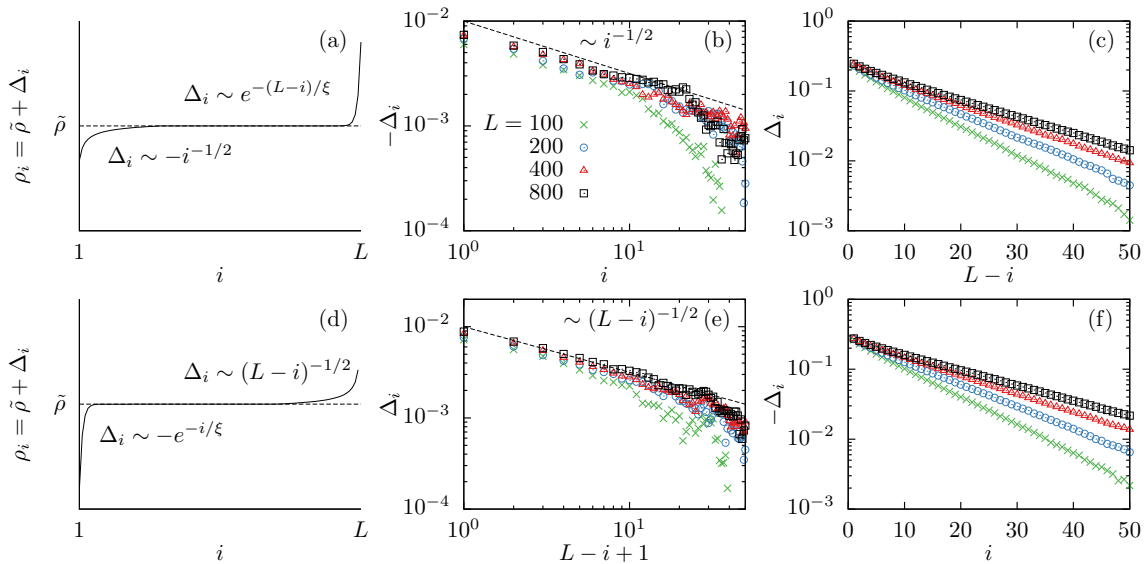}
    \caption{\label{fig:density_profile} (Color online) (a) Density profile of particles in the low-density phase. For $\rho = 0.34$, the particle density converges to the bulk value $\tilde{\rho}$ (b) algebraically near the entrance and (c) exponentially near the exit. (d) Density profile of particles in the high-density phase. For $\rho = 0.66$, the particle density converges to the bulk value $\tilde{\rho}$ (e) algebraically near the exit and (f) exponentially near the entrance. The CDRR networks with $N = 10^3$ and $c = 2$ are used.}
\end{figure*}

The density--current relation obtained by the DMF and the DPMF
theories are compared with the simulation results obtained at
different values of the segment length $L$ in
Fig.~\ref{fig:den_cur}. As $L$ increases, the current plateau
becomes narrower in the $\rho$-axis and higher in the $J$ axis
than predicted by the DMF theory, which qualitatively agrees with
the corrections predicted by our DPMF argument. The inset of
Fig.~\ref{fig:den_cur}, which shows the covariance between each
junction and its adjacent sites defined by
\begin{align}
\sigma_{Lu} &\equiv \langle \tau_L \tau_u \rangle - \langle \tau_L \rangle \langle \tau_u \rangle, \nonumber \\
\sigma_{u1} &\equiv \langle \tau_u \tau_1 \rangle - \langle \tau_u \rangle \langle \tau_1 \rangle,
\end{align}
also shows that the junctional correlations are generally non-negligible.

We also check the accuracies of the theories by observing how the deviations of $J$ from the predictions of each theory scale with $L$. For the CDRR networks with $c = 2$, the inset of Fig.~\ref{fig:scaling} shows that deviations from the DMF predictions constantly decrease with $L$ only for $\rho = 0.3$ and $0.7$. The DMF theory obviously fails to predict the asymptotic values of $J$ for the other values of $\rho$, all of which are supposed to be in the phase coexistence regime according to the theory. Meanwhile, the main plot of Fig.~\ref{fig:scaling} shows that the DPMF predictions are generally more accurate than the DMF counterparts, although the extent of accuracy in the asymptotic limit remains to be estimated by increasing $L$ even further. If the DPMF predictions are indeed accurate for some values of $\rho$, then the finite-size corrections seem to scale as $L^{-1}$. This might be because the junctions create a difference between the finite-size bulk density $\tilde{\rho}$ and the global density $\rho$, which satisfies $\tilde{\rho} - \rho \sim L^{-1}$.

\subsection{Homogeneity of the steady state}

The DMF and the DPMF theories both predict that the TASEP on the CDRR networks has a homogeneous steady state, in which all junctions have the same mean occupancy, and all links have the same current. In order to check this prediction, we define the average of a junction (link) observable $X_u$ ($X_{uv}$) over different junctions (links) of the system as
\begin{align}
\overline{X_u} \equiv \frac{1}{N} \sum_u X_u, \quad \overline{X_{uv}} \equiv \frac{1}{cN} \sum_{uv} X_{uv}.
\end{align}
If the prediction is true, then we must have
\begin{equation} \label{eq:homogeneity}
\overline{\rho_u^2} - \overline{\rho_u}^2 = \overline{J_{uv}^2} - \overline{J_{uv}}^2 = 0.
\end{equation}

The above condition is checked in Fig.~\ref{fig:time_variance}, which shows the estimations of $\overline{\rho_u^2} - \overline{\rho_u}^2$ and $\overline{J_{uv}^2} - \overline{J_{uv}}^2$ in a single CDRR network with $N = 10^3$, $c = 2$, and $L = 10^2$ as functions of the observation time $T$ for different values of $\rho$. Here $\rho_u$ of each junction and $J_{uv}$ of each link are estimated as time averages over a period of length $T$ rather than ensemble averages. If $T$ is sufficiently greater than the autocorrelation time, then we expect $T$ to be proportional to the effective number of uncorrelated samples, making the time averages equivalent to the ensemble averages in the limit $T \to \infty$. Indeed, Fig.~\ref{fig:time_variance} shows that the variances asymptotically decrease as $T^{-1}$ in all cases. This behavior strongly suggests that the variances are not due to any essential differences between junctions or links but due to the finite number of uncorrelated samples. Thus, it seems that Eq.~(\ref{eq:homogeneity}) is true for any realization of a CDRR network, even if the network has a finite number of sites.

It is notable that the estimated variance of $\rho_u$ in the phase coexistence regime ($\rho = 0.5$) exhibits a crossover from the $T^{-2/3}$ decay to the $T^{-1}$ decay. When low-density and high-density blocks coexist in each link, the interface between them prevents the spread of local perturbations~\cite{Kolomeisky1998b}, lengthening the autocorrelation time of the system. If $T$ is not sufficiently larger than the autocorrelation time, then the time dependence of the estimated variance of $\rho_u$ reflects the relaxation of local perturbations. The observed $T^{-2/3}$ decay is consistent with the fact that the length scale of a local perturbation grows with time as $t^{2/3}$ for systems described by the noisy Burgers equation~\cite{vanBeijeren1985} or the Kardar-Parisi-Zhang (KPZ) surface growth dynamics~\cite{Kardar1986}, of which the TASEP is one example.

\subsection{Density profiles}

Figure~\ref{fig:density_profile} shows the steady-state density profile averaged over all links of the system, where $\Delta_i \equiv \rho_i - \tilde{\rho}$ denotes the difference between the mean occupancy of the $i$-th inner site and the effective bulk density $\tilde{\rho}$ estimated by the relation $J = \tilde{\rho}(1-\tilde{\rho})$.

The upper panel of Fig.~\ref{fig:density_profile} shows the density profile of particles in the low-density phase with $\rho = 0.34$. Near the entrance, the particle density increases algebraically to the bulk density with the exponent close to $1/2$, which can be written as
\begin{equation} \label{eq:LD_profile_ent}
\Delta_i \sim i^{-1/2}.
\end{equation}
Meanwhile, near the exit, the particle density decreases exponentially to the bulk density as
\begin{equation} \label{eq:LD_profile_ext}
\Delta_i \sim \exp \left[-(L-i)/\xi \right].
\end{equation}
While Eq.~(\ref{eq:LD_profile_ext}) is qualitatively consistent with the 1D result, Eq.~(\ref{eq:LD_profile_ent}) is not. The latter is against the property of the open 1D system that the algebraic decay with the exponent $1/2$ appears only in the maximal-current phase~\cite{Blythe2007}.

Similar difference from the homogeneous 1D TASEP is also observed in the high-density phase, as shown in the lower panel of Fig.~\ref{fig:density_profile} for $\rho = 0.66$. In this case, the density profile has an inverted shape so that
\begin{equation} \label{eq:HD_profile_ext}
\Delta_i \sim (L-i)^{-1/2}
\end{equation}
near the exit and
\begin{equation} \label{eq:HD_profile_ent}
\Delta_i \sim \exp \left(-i/\xi\right)
\end{equation}
near the entrance. Again, Eq.~(\ref{eq:HD_profile_ext}) shows a behavior that cannot occur outside the maximal-current phase. These properties of the density profiles indicate that the TASEP on the CDRR networks has anomalies that cannot be explained by any analytical approaches relying on the effective rate assumptions.

\section{Slow-bond effects} \label{sec:SB}

The algebraic behavior of density profiles outside the maximal-current phase was previously reported in the context of the 1D TASEP with open boundaries and a single slow bond of hopping rate $r$ in the middle~\cite{MHa2003}. The study found that the particle density decays to the bulk values algebraically with exponent $1/2$ before and after the slow bond if $r < r_c \approx 0.80(2)$, which agrees with our observation presented in the previous section.

What is the origin of this agreement? In the model defined in Sec.~\ref{sec:model}, every site of the system rather than every bond is updated at an equal rate. Both update schemes are equivalent for 1D systems, where every site has the same number of bonds attached to it. However, if the bonds are unevenly distributed among sites and if the system is updated sitewise, then the bonds attached to the sites with high connectivity become effectively slow. This is indeed what happens to the bond between the entrance junction and the first inner site of our model. Because of these slow bonds, many properties of the 1D slow-bond problem are still retained in the network case.

The slow-bond effects explain the deviations from the DMF theory other than the algebraic boundary behavior. First, the particle--hole asymmetry observed at finite size (asymmetry with respect to the $\rho \to 1-\rho$ transformation observed in Figs.~\ref{fig:den_cur} and \ref{fig:scaling}) originates from the asymmetric arrangement of slow bonds. Since the particles have a slow bond near the entrance and the holes have their own near the exit, the particle--hole symmetry is violated on a microscopic level. Second, the correlations at the entrance and exit junctions (see the inset of Fig.~\ref{fig:den_cur}) show the influence of the slow bonds. The covariance between the entrance junction and the first inner site measures the correlation between the endpoints of a slow bond, which is naturally bound to be negative. On the other hand, the covariance between the exit junction and the last inner site changes sign depending on the direction of information flow near the exit, which is determined by the group velocity $v_g = 1 - 2\tilde{\rho}_\mathrm{exit}$ of density waves~\cite{Kolomeisky1998b}. Here $\tilde{\rho}_\mathrm{exit}$ denotes the effective bulk density near the exit. When the information is flowing from the bulk (low-density phase near the exit), the covariance becomes positive due to the accumulation of particles before the slow bond. On the other hand, when the information is flowing from the nearest slow bonds (high-density phase near the exit), the covariance becomes negative due to their influence.

These connections to the slow-bond problem also have interesting implications for the effects of junction regulations. A recent numerical study about the TASEP on closed loops sharing a single junction~\cite{Raguin2013} showed that the range of the coexistence regime can be reduced by pumping the particles out of junctions. It was observed that if the junctions are updated $c$ times as fast as the ordinary sites, then the coexistence regime completely disappears, so that the maximal current $J = 1/4$ can be achieved. Such behavior makes sense because the pumping rate of $c$ eliminates the effective slow bonds next to junctions.

However, is $c$ the {\em minimal} pumping rate for the elimination of the current plateau? This question is related to the unresolved problem of the critical hopping rate $r_c$ of the slow bond, above which the slow bond cannot create a macroscopic traffic congestion. Some studies~\cite{MHa2003,JHLee2009} report $r_c < 1$, while others support $r_c = 1$~\cite{Janowsky1992,*Janowsky1994,Kolomeisky1998a,Tang1993,*Balents1994,*Krug1994,*Kinzelbach1995,*Hwa1995,*Lassig1998,Costin2012}. If the former is true, then the minimal pumping rate required for the maximal current might be lower than $c$, which would be another important consequence of the junctional correlations.

\section{Conclusions}
\label{sec:conclusions}

We have discussed the effects of correlations between junctions
and their adjacent sites in the TASEP on the closed and directed
random regular networks. Our defect pair mean-field theory showed
that the range of the phase coexistence regime must be narrower
and that the height of the current plateau must be higher than the
simple mean-field predictions neglecting the junctional
correlations. The numerically obtained fundamental diagram (i.e.,
density--current relation) and the scaling behaviors of the
steady-state current confirmed the existence of those corrections.
Moreover, we also observed the algebraic convergence of the
density profiles to bulk density values with exponent close to
$1/2$, which is attributable to the slow-bond effects.
Implications of those slow-bond effects for the TASEP in more
heterogeneous network structures and the optimal junction
regulations remain interesting subjects for future studies.

\begin{acknowledgements}
This work was supported by the National Research Foundation of
Korea (NRF) funded by the Korea government [Grant No.
2014R1A1A4A01003864 (Y.B., M.H.) and No. 2011-0028908 (Y.B.,
H.J.)].
\end{acknowledgements}

\bibliography{PRE2014-EX11081-final}

\end{document}